\documentclass[12pt]{article}
\usepackage{graphicx}
\begin{document}
%
%
\begin{center}
{\large  The $\pi \rightarrow \pi\pi$ process in nuclei
and the restoration of chiral symmetry}
\end{center}
%
%
\begin{center}
{\large $CHAOS$  Collaboration}
\end{center}
%
%
\begin{center}
\normalsize
N. Grion$^{a,}$\footnote[1]
{Corresponding author, electronic mail: Nevio.Grion@ts.INFN.it}, 
M. Bregant$^{a,b}$, P. Camerini$^{a,b}$, E. Fragiacomo$^{a,b}$, 
S. Piano$^{a,b}$, R. Rui$^{a,b}$, E.F. Gibson$^{c}$, G. Hofman$^{g}$
E.L. Mathie$^{d}$, R. Meier$^{e}$, M.E. Sevior$^{f}$, 
G.R. Smith$^{g,}$\footnote[2] 
{Present address: Jefferson Lab, Newport News, Va 23006, USA}
and R. Tacik$^{d}$.
\end{center}
%
%

\begin{center}
\small{\it 
$^a$ Istituto Nazionale di Fisica Nucleare, 34127 Trieste, Italy \\
$^b$ Dipartimento di Fisica dell'Universita' di Trieste, 34127 Trieste, 
     Italy\\
$^c$ California State University, Sacramento CA 95819, USA \\
$^d$ University of Regina, Regina, Saskatchewan, Canada S4S 0A2 \\
$^e$ Physikalisches Institut, Universit\"{a}t T\"{u}bingen, 
     72076 T\"{u}bingen, Germany \\
$^f$ School of Physics, University of Melbourne, Parkville, Vic., 3052,
     Australia \\
$^g$ TRIUMF, Vancouver, B.C., Canada V6T 2A3 \\
} 
\end{center}                   

\setlength{\baselineskip}{2.3ex}         
{\small\bf Abstract:}
{\small The results of an extensive campaign of measurements of the  
$\pi\rightarrow\pi\pi$ process in the nucleon and nuclei at intermediate 
energies are presented. The measurements were motivated by the study of 
strong $\pi\pi$ correlations in nuclei. The analysis relies on the 
composite ratio $\cal C$$_{\pi\pi}^A$, which accounts for the clear 
effect of the nuclear medium on the $\pi\pi$ system. The comparison of 
the $\cal C$$_{\pi\pi}^A$ distributions for the  $(\pi\pi)_{I=J=0}$ and    
$(\pi\pi)_{I=0,J=2}$ systems to the model predictions indicates that the 
$\cal C$$_{\pi\pi}^A$ behavior in proximity of the 2m$_\pi$ threshold is 
explainable through the partial restoration of chiral symmetry in nuclei. 
}

\normalsize
%

\begin{center}
{\bf I. INTRODUCTION}
\end{center}

Spectral properties of pion pairs interacting in the I=J=0 channel (the 
$\sigma$-channel) are predicted to vary significantly from the vacuum 
to nuclear matter as a consequence of the partial restoration of chiral 
symmetry. As an example, the vacuum spectral function of $\sigma$, a 
broad ($\Gamma\sim$500 MeV) resonance centered at $\sim$500 MeV, 
substantially reshapes in nuclear matter by forming a peak-like
structure at around 2m$_\pi$ \cite{theory:one,theory:two,theory:three}. 
The underlying theory regards the $\sigma$ meson as a $\bar{q}q$ 
excitation of the QCD vacuum, in which the spontaneous breaking of the 
chiral symmetry leads to the $\sigma$-$\pi$ mass difference. The sigma 
(J$^P=0^+$) is also the chiral partner of the pion (J$^P=0^-$). When 
the properties of the $\sigma$ meson are studied in nuclear matter, 
the theory predicts a substantial change  of the $\sigma$ spectral 
function, which strongly reduces the $\sigma$-$\pi$ mass difference. 
This occurrence indicates that nuclear matter partially restores the 
chiral symmetry.  The I=0 $\pi\pi$ interaction in nuclear matter 
is also studied in Ref.\cite{theory:three.five}, which reflects 
the current theoretical understanding on this topic.

An additional source of reshaping of the $\sigma$ spectral function at 
around threshold is yielded by standard many-body correlations; i. e.,  
the $P-$wave coupling of pions to $particle-hole$ and $\Delta-hole$ 
states \cite{theory:two,theory:three,theory:four}. The combined effect 
of partial restoration and collective P-wave pionic modes produces a 
conspicuous enhancement  of the $\sigma$ spectral function at around 
the 2m$_\pi$ threshold \cite{theory:two,theory:three}. This letter 
presents further analysis of experimental results on the 
$\pi\rightarrow\pi\pi$ process near the 2m$_\pi$ threshold, which are 
then related to the direct observation of $\pi\pi$ in-medium 
correlations. In this regard, final pion pairs are studied in the 
vacuum and in the nuclear medium, and are further examined in the 
isospin 0 and 2 channels. The comparison of different isospin channels 
conveys additional information on the spectral changes of the 
$\sigma$-channel (I=0) with respect to the non-resonant I=2 channel. 
Finally, the data from the present measurements will directly probe 
the $\sigma$-spectral predictions around threshold and accordingly 
the underlying physics of chiral symmetry restoration.

The $\sigma$ (or $f_0$(600)) meson is understood to be a broad 
resonant state $\Gamma_\sigma\sim m_\sigma\sim $500 MeV  which 
predominantly decays into two S-wave pions $\sigma \rightarrow \pi\pi$ 
\cite{pdg:one}. The $\sigma$  broad structure makes this meson 
difficult to directly observe via the $\pi N\rightarrow\pi\pi N$ 
elementary reaction \cite{expt:one}, or heavy meson decays 
\cite{expt:two}. A systematic analysis of a broad sample of data 
involving pion pairs in the I=J=0 channel however provides firm 
evidence of $\sigma$ \cite{theory:five}. A clear signature of 
$\sigma$ in the vacuum appears controversial. Conversely, the nuclear 
medium may condensate I=0 pion pairs by changing the structure of the 
QCD vacuum; therefore, the study of $\sigma$ by means of two coincident 
I=0 pions via the $\pi\rightarrow\pi\pi$ process appears appropriate. 

The $\sigma$ spectral properties are studied by means of the $\pi\pi$ 
invariant mass and the composite observable $\cal C$$_{\pi\pi}^A$, 
which is described in Sec. 3. In order to normalize this observable to 
pion production on the nucleon and explicitly consider the ratio for 
nuclei from $^2$H to  $^{208}$Pb, a  new analysis of our previously 
published \cite{expt:twopointfive} pion production data on the nucleon 
was completed as a function of the same kinematic quantities as were 
used for the nuclear data.  $\cal C$$_{\pi\pi}^A$ appears slightly 
different from the previously published one, which was normalized to 
deuterium \cite{expt:three}. In addition, new results for the composite 
observable are presented for Sc as a function of incident energy.  The 
final pions have an energy distribution which is broadly centered 
between 20-50  MeV, depending on the energy of the projectile 
\cite{expt:three,expt:four}.  In this energy range, an earlier 
$\pi 2\pi$ measurement reported that the shape of spectra was only
slightly altered by final state interactions \cite{expt:five}. This 
result finds a qualitative explanation in the long mean-free path of 
low-energy pions in nuclear matter. The mean-free path of pions exceeds 
10 fm for $\rho = \rho_0/2$ at T$_\pi$=50 MeV \cite{theory:six}, which 
highly reduces the pion distortions due to $\pi$N final-state 
interactions.  Furthermore, pion absorption has only the effect of 
removing final pions and the removal is mildly dependent on the pion 
energy in the interval 50$\pm$30 MeV \cite{theory:seven} thus causing 
little reshaping of pion spectra. Finally, pion distortions due to other 
$\pi$A reactions can safely  be neglected \cite{theory:seven}. Previous 
studies of the $\pi A\rightarrow \pi\pi A^\prime$ reaction showed that 
at intermediate energies the pion production process takes place at the 
nucleus skin  $\rho\sim\frac{1}{3}\rho_0$, where $\rho_0$ is the nuclear 
density at saturation \cite{expt:four}. This occurrence further increases 
the nuclear transparency to final pion pairs. The moderate fraction of 
$\rho_0$ inspected by pions however limits the $\sigma$ spectral changes. 
These can only be augmented by probing higher nuclear densities, which 
can be obtained with the use of electromagnetic beams. This in fact was 
the approach used by the authors of Ref. \cite{expt:six}, who were able 
to probe nuclear densities up to $\rho\sim\frac{2}{3}\rho_0$ for 
$^{208}Pb$ by employing $\gamma$'s of energies in the range 400-460 MeV. 
     
\begin{center}
{\bf II. THE MEASUREMENT}
\end{center}

The data were taken at the M11 pion channel of TRIUMF with the CHAOS 
spectrometer \cite{CHAOS:one}. Both positive and negative pions 
were used to study the pion-production reaction ($\pi2\pi$) as a 
function of the atomic mass number (A) and the projectile incident 
energy (T): 
\begin{equation}
        \pi^{\pm} p \rightarrow \pi^+\pi^{\pm} n \textrm{ at } 
        T=\textrm{243, 264, 284  and  305} \textrm{ MeV }
\end{equation}
\begin{equation}
        \pi^+ A \rightarrow \pi^+\pi^{\pm} X  \textrm{ with }
        A\textrm{: } ^{2}H, ^{12}C, ^{40}Ca\textrm{ and }^{208}Pb 
        \textrm{ at }T=283 \textrm{ MeV }
\end{equation} 
\begin{equation}
        \pi^+ Sc \rightarrow \pi^+\pi^{\pm} X \textrm{ at }
        T=\textrm{243, 264, 284 and 305} \textrm{ MeV }
\end{equation}
The measurements from (1) to (3) were performed under the same 
kinematic conditions to ensure a direct comparison among the 
$\pi 2\pi$ data. In addition, final pion pairs were detected in 
coincidence to avoid the overwhelming background from the reaction 
of pion scattering.  The results of the measurements (1), (2) 
and (3) were previously published in Refs.\cite{expt:twopointfive}, 
\cite{expt:three} and \cite{expt:four}, respectively.  

The CHAOS spectrometer consists of a dipole magnet, four rings of 
cylindrical wire chambers  and a multilayer mass-identification 
system (CFT). The events were analyzed  in the plane of the reaction 
due to the geometry of the dipole; that is, for azimuth angles $\Theta$ 
ranging from 0$^\circ$ to 360$^\circ$ and for zenith angles $\Phi$ from 
-7$^\circ$ to +7$^\circ$. Final pions (from (1) to (3)) were detected 
with an energy resolution of $\delta$T/T=4-7\% ($\sigma$), for the 
CHAOS field set at 0.5-0.6 T. Such a field also established the pion 
detection threshold, which was 11 MeV. The CFT system was designed to 
delivered the first level trigger and mass identify charged particles; 
i.e., $e$'s, $\pi$'s, $p$'s and $d$'s. The particle mass identification 
(PID) relies on the observed correlation between the trace momentum and 
the pulse heights in the CFT layers. Reconstructed events with a valid 
PID were further restricted to the $\pi 2\pi$ phase-space volume before 
being finally saved. The prime PID capability of CHAOS is illustrated 
in Fig. 1, which shows the missing mass distribution of reactions (1) 
at T=264 MeV. The missing mass expected is the neutron mass irrespective 
of the projectile charge and energy. In fact, a distinct  peak centered 
around 940 MeV with a $\sigma$ of about 3 MeV is observed in each channel. 
By averaging the peak-value over the four energies and the two reaction 
channels, the missing mass distribution yields a mean value of 
941.2$\pm$3.0 MeV which is consistent with the neutron mass.  
%
%
\begin{figure}[t,c,b]
 \centering
  \includegraphics*[angle=90,width=0.7\textwidth]
                   {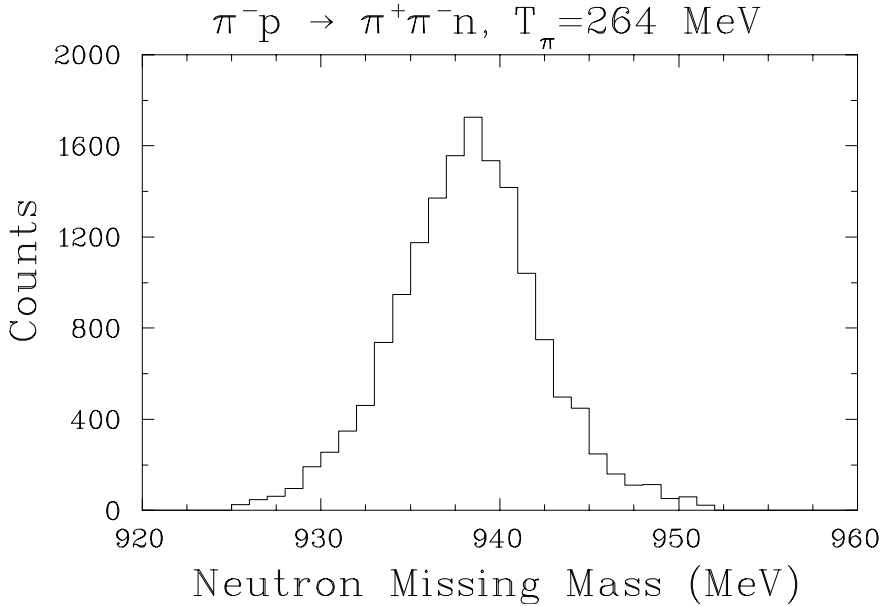}
    \caption{\footnotesize  Missing mass distribution of the  
            $\pi^- p\rightarrow\pi^+\pi^- n$ reaction at an incident 
            projectile energy of 264 MeV. The expected missing mass is 
            the neutron mass, the distribution is peaked at 938.3 MeV.}
\end{figure}      

\begin{center}
\bf {III. THE OBSERVABLE}
\end{center}

The $\pi 2\pi$ experimental results will be compared to the theoretical 
predictions via the composite observable 
\begin{equation}
    \cal C _{\pi\pi}^\mathrm{A} =  \frac
   {\sigma(\mathrm{M}_{\pi\pi}^\mathrm{A}) / \sigma_\mathrm{T}^\mathrm{A}}
   {\sigma(\mathrm{M}_{\pi\pi}^\mathrm{N}) / \sigma_\mathrm{T}^\mathrm{N}}
\end{equation}
where $\sigma(M_{\pi\pi}^A)$ ($\sigma(M_{\pi\pi}^N)$) is the triple 
differential cross section 
$d^3\sigma/dM_{\pi\pi}d\Omega_\pi d\Omega_\pi$ for nuclei (nucleon), 
$M_{\pi\pi}$ represents the $\pi\pi$ invariant mass, $\Omega_\pi$ 
denotes the solid angle into which a charged pion is scattered, and 
$\sigma_T^A$ ($\sigma_T^N$) is the total cross section in nuclei 
(nucleon). Differential as well as total cross sections were 
determined by means of the CHAOS measurements, details of the 
experimental method used are reported in \cite{expt:three,expt:four}.
 The  extrapolation of the raw data to the full solid angle is 
model-dependent because of the limited  $\Phi$-acceptance of the 
magnetic spectrometer. A detailed discussion is reported in Ref. [12]. 
The effects of such an extrapolation must be accounted for when 
comparing the CHAOS cross sections to model predictions. However, 
these effects cancel to a large extent in the composite ratio 
$C_{\pi\pi}^{A}$, which therefore is better suited for the discussion 
of the observed effects than the invariant mass distributions themselves.

Such an observable is useful to focus on the medium modification of 
meson properties. $\cal C$$_{\pi\pi}^A$ in fact describes the clear 
effects of the nuclear medium on the $\pi\pi$ interacting system.
The pion-production reaction in nuclei is a quasi-free process, which 
requires a single  nucleon $\pi N \rightarrow \pi\pi N$ 
\cite{expt:five}; therefore, the ratio of $\sigma(M_{\pi\pi}^{A})$ 
to $\sigma(M_{\pi\pi}^{N})$  is loosed from the reaction mechanism, 
and accordingly is $\cal C$$_{\pi\pi}^A$. The normalization of
$\sigma(M_{\pi\pi}^{A})$ to $\sigma_T^A$ removes the dependence of  
$\cal C$$_{\pi\pi}^A$ from the number of scattering centers in nuclei, 
since both terms depend equally on $A$. Furthermore, the limited 
acceptance of CHAOS should slightly affect $\cal C$$_{\pi\pi}^A$ 
since the detector acceptance is the same for N and A. In order to 
verify such an assumption, the behavior of $\cal C$$_{\pi\pi}^A$ 
was simulated for CHAOS (open histogram) and an ideal $4\pi$ detector 
(filled histogram), and the results of the simulations are shown 
in Fig. 2 for the $\pi^+$ $^{45}Sc \rightarrow \pi^+ \pi^-p$ $^{44}Sc$ 
(upper panel) and $\pi^+$ $^{45}Sc \rightarrow \pi^+ \pi^+n$ $^{44}Ca$ 
(lower panel) reactions at 284 MeV. 
%
%
\begin{figure}[t,c,b]
 \centering

  \includegraphics*[angle=90,width=0.7\textwidth]
                   {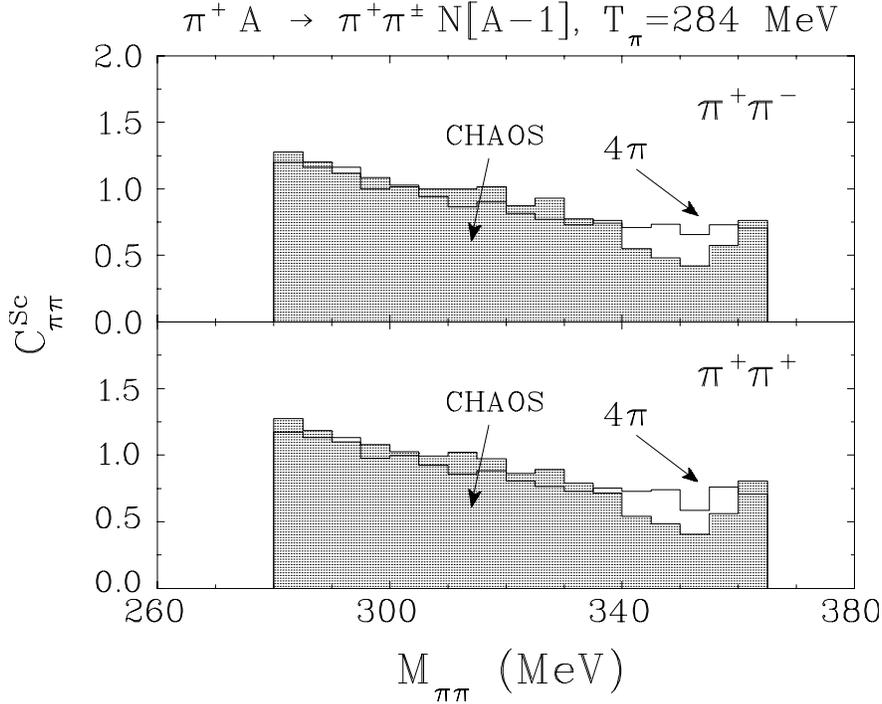}
    \caption{\footnotesize Simulated behavior of $\cal C$$_{\pi\pi}^{Sc}$ 
             for CHAOS (filled histogram) and for an ideal $4\pi$ detector 
             (open histogram) as a function of the $\pi\pi$ invariant mass 
             (M$_{\pi\pi}$). The reactions simulated are 
             $\pi^+$ $^{45}Sc \rightarrow \pi^+\pi^- p$ $ ^{44}Sc$ 
             (upper panel) and  
             $\pi^+$ $^{45}Sc \rightarrow \pi^+\pi^+ n$ $ ^{44}Ca$  
             (lower panel) at an incident projectile energy 
             of 284 MeV.}
\end{figure}
The histograms were plotted without  requiring any normalization, but the 
observables forming the nominator (denominator) of $\cal C$$_{\pi\pi}^A$ 
were generated by feeding the Monte Carlo code with the same number of 
input events.  The out-of-plane behavior of the 
$\pi N (A) \rightarrow \pi \pi N (A')$ reactions was accounted for by 
the model described in Ref.\cite{theory:four} (and references therein 
quoted).  The histograms display a monotonic decrease and nearly the 
same intensities at the varying of $M_{\pi\pi}$, with the exception of 
a shallow dip at $M_{\pi\pi} \sim$350 MeV for the CHAOS distribution. 
This convincingly demonstrates that $\cal C$$_{\pi\pi}^A$ is both 
weakly related to the detector acceptance and  nearly independent of 
the reaction channel.      
      
Some models which describe the $\pi\pi$ dynamics in nuclear matter do 
not deal with the $\pi 2\pi$ reaction mechanism nor account for the 
nuclear structure \cite{theory:one,theory:two,theory:three}. These 
models focus on  understanding how nuclear matter alters the vacuum 
structure of QCD and the repercussions on the spectral properties of 
mesons and hadrons. Mesons are of prime interest because they are 
considered as the elementary $\bar{q}q$ excitation of the vacuum. 
In this framework, the observable $\cal C$$_{\pi\pi}^A$ can be 
quantitatively compared to the model predictions.  At the variance, 
the model quoted in Ref.\cite{theory:four}, provides a comprehensive 
study of the $\pi A \rightarrow \pi\pi X$ reaction: it accounts for the 
elementary process of pion production $\pi N \rightarrow \pi\pi N$ as 
well as standard nuclear effects (Pauli blocking, Fermi motion, etc.). 
It also examines the effects of the nuclear environment on J=I=0 
interacting pion pairs  via the $P-$wave coupling of pions to 
$particle-hole$ and $\Delta-hole$ configurations \cite{theory:nine}. 
The model, which embeds the CHAOS acceptance, predicts 
$\sigma(M_{\pi\pi}^{A(N)})$ and $\sigma_T^{A(N)}$ therefore  
$\cal C$$_{\pi\pi}^A$ for the $\pi^+ \rightarrow\pi^+\pi^\pm$ 
reaction channels.
      
\begin{center}
{\bf IV. THE  A- AND T-DEPENDENCE OF   $\cal C$$_{\pi\pi}^A$}
\end{center}

The error bars of the $\cal C$$_{\pi\pi}^A$ data points plotted in 
figures from 3 to 5 account solely for statistical uncertainties,
%
%
\begin{figure}[b,c,t]
 \centering
  \includegraphics*[angle=90,width=1.0\textwidth]
                   {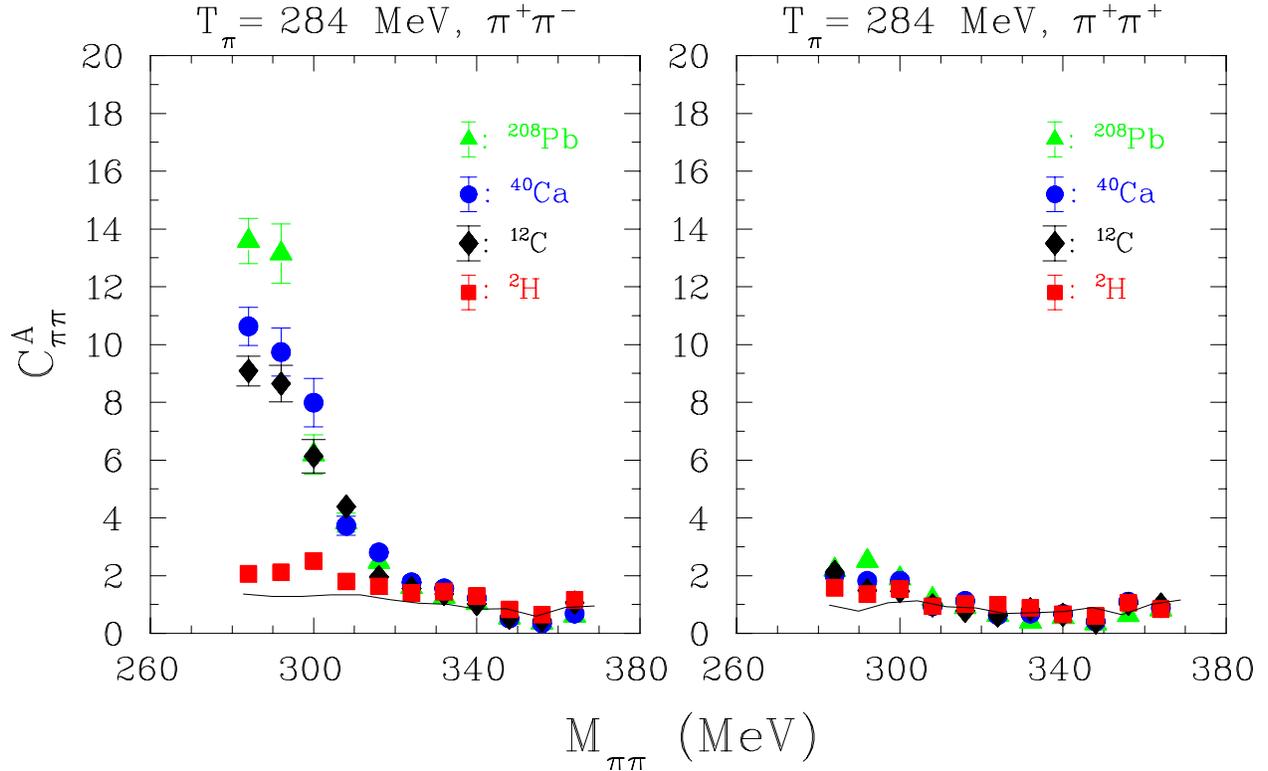}
    \caption{\footnotesize A-dependence of  $\cal C$$_{\pi\pi}^{A}$ as a 
             function of the  $\pi\pi$ invariant mass at an incident 
             projectile energy of 284 MeV. The continuous lines are the 
             result of the model predictions of Ref.\cite{theory:four}.}
\end{figure}
which primarily reflect those of measurements (2) and (3). The 
systematic uncertainties associated to $\cal C$$_{\pi\pi}^A$ range 
from 15.0\% (A-dependence, measurement (2)) to 16.7\% (T-dependence, 
measurement (3)),which must be summed 
in quadrature with the statistical ones to obtain the overall 
uncertainties. The A-dependence of the composite ratio is reported 
in Fig. 3 for a projectile kinetic energy of 284 MeV. In the 
$\pi^+\pi^-$ channel, pion pairs largely  ($\sim$95\%) couple 
to I=J=0 quantum numbers \cite{expt:three,theory:ten};  however, 
the fraction of I=J=0 pion pairs which couple to the $\sigma$-meson 
cannot be established by the present measurement.  The pure 
isospin I=2 is instead always reached  by $\pi^+\pi^+$ pairs. 
In this isospin  channel, $\cal C$$_{\pi\pi}^A$ barely depends on 
A: data overlap from $^{2}H$ to $^{208}Pb$ and the weak threshold 
enhancement is primarily due to phase space as denoted by the 
$\cal C$$_{\pi\pi}^{Sc}$ behavior in Fig. 2 (see also discussion 
of Fig. 5). The $\sigma$-channel is characterized by a substantial 
dependence on A but only for $A>2$, when nuclear matter is realized. 
The $\cal C$$_{\pi\pi}^{A}$ observable is compared with the model 
predictions of Ref.\cite{theory:four} for A=$^{40}Ca$ (continuous 
line). The agreement is good for the $\pi^+\pi^+$ channel,  
except for the low-energy part of the spectrum where the model
calculations underestimate the data by nearly a factor of two.  
The model fails also to reproduce the threshold  enhancement of 
$\cal C$$_{\pi\pi}^{A}$ for the $\pi^+\pi^-$ channel, which 
indicates that the model fails to account for medium modification 
on pion pairs interacting in the $\sigma$-channel. In the framework 
of this model, an explanation of the threshold enhancement of 
$\cal C$$_{\pi\pi}^{A}$ may be related to the modifications of the 
$\pi \rightarrow \pi \pi$ elementary amplitude inside nuclear 
matter.  In fact, a modification of some pieces of this amplitude 
may modify the strong interferences present at threshold, thus 
causing  a  significant  reshaping of  $\cal C$$_{\pi\pi}^{A}$. 
Studies are in progress \cite{theory:eleven}.  
%
%
\begin{figure}[t,c,b]
 \centering
  \includegraphics*[angle=90,width=1.0\textwidth]
                   {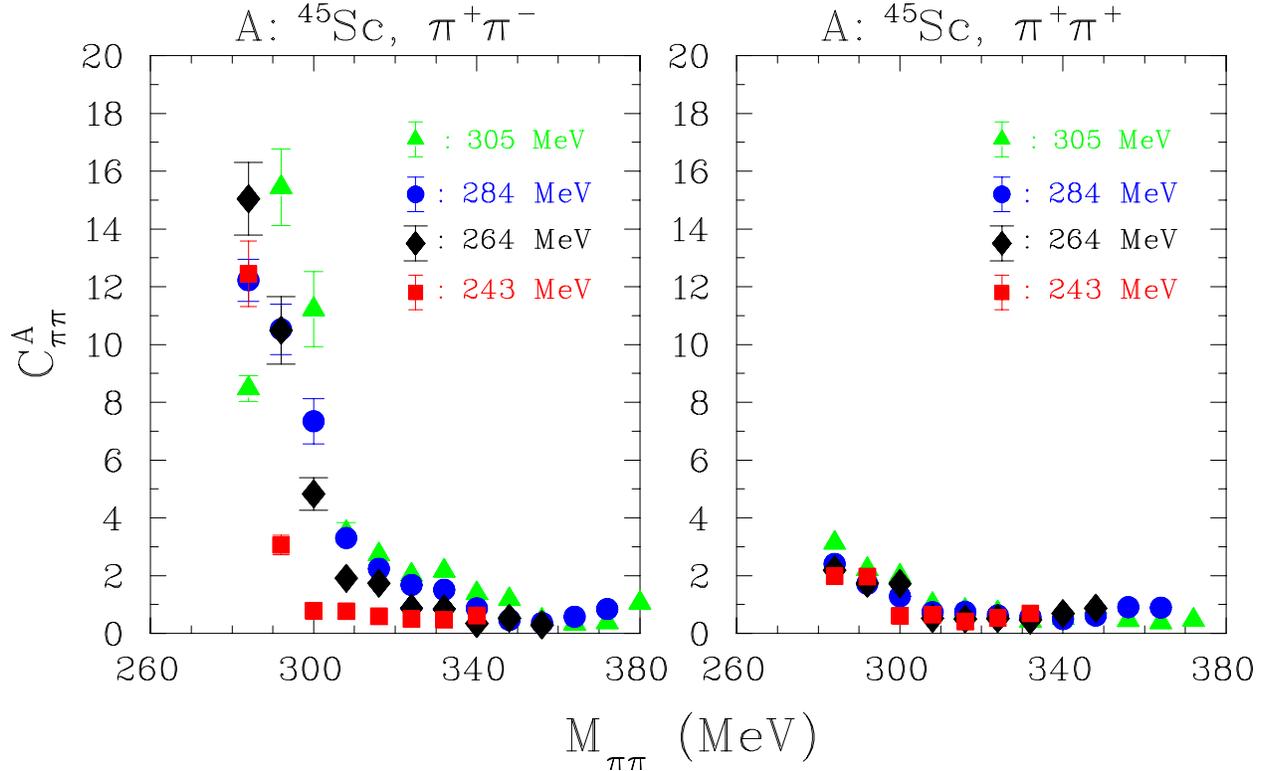}
    \caption{\footnotesize T-dependence of  $\cal C$$_{\pi\pi}^{A}$ as 
             a function of the  $\pi\pi$ invariant mass.}
\end{figure}
The T-dependence study of $\cal C$$_{\pi\pi}^A$ done for 
A=$^{45}Sc$ delivers the same general picture as the A-dependence, 
Fig. 4. A strong enhancement is observed for $\cal C$$_{\pi\pi}^{A}$ 
in proximity of the 2m$_\pi$ threshold for the isospin 0 channel 
(left panel), while $\cal C$$_{\pi\pi}^{A}$ is flat over the $\pi\pi$ 
invariant mass range for I=2 (right panel). For a selected isospin 
channel, $\cal C$$_{\pi\pi}^{A}$ depicts a behavior weakly varying 
with T. This is consistent with a previous study on the properties 
of the pion production reaction, which were based on the model 
calculation of Ref.\cite{theory:four}. The study shows that the 
average nuclear density ($\rho$) probed by incident pions barely 
changes from 240 to 320 MeV. In fact, $\rho \sim 0.36\rho_0$, which 
localizes the reaction to occur at the nucleus surface.   

In order to compare the $\cal C$$_{\pi\pi}^A$  distributions with 
similar results from other available theories 
\cite{theory:one,theory:two,theory:three}, the $\pi 2\pi$ data must 
be normalized to their phase space. This is because such theories deal 
only with pion pairs in nuclear matter taking no regard to the reaction 
of pion production; therefore, to its phase space. Fig 5 depicts the 
%
%
\begin{figure}[t,c,b]
 \centering
  \includegraphics*[angle=90,width=0.9\textwidth]
                   {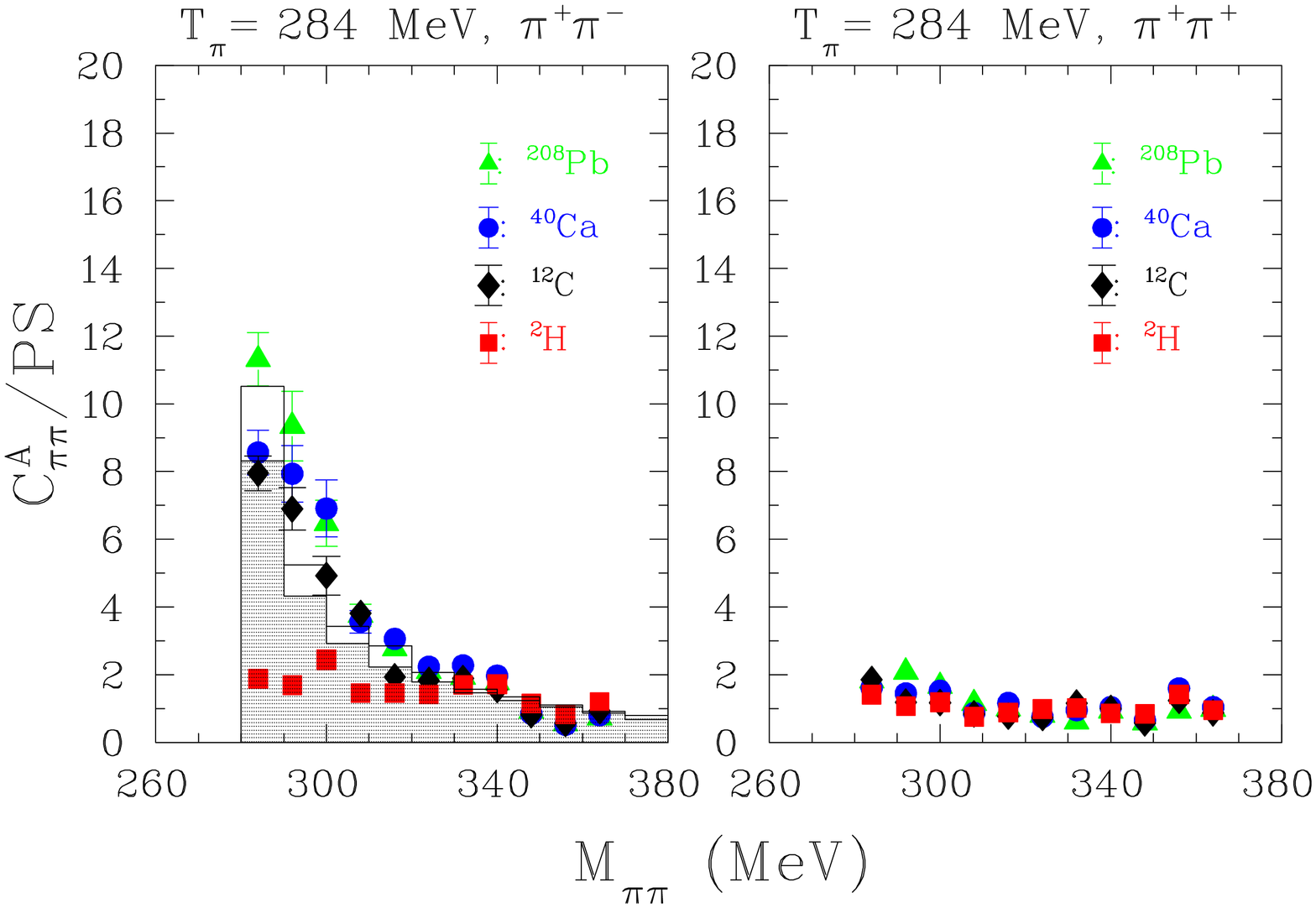}
    \caption{\footnotesize A-dependence of $\cal C$$_{\pi\pi}^{A}$ 
             normalized to the $\pi A \rightarrow \pi\pi N [A-1]$ phase 
             space (PS) at an incident projectile energy of 284 MeV. 
             The phase-space simulations embody the CHAOS acceptance. 
             The open  and filled diagrams are the result of the model 
             predictions of \cite{theory:three} and \cite{theory:one}, 
             respectively, which are normalized to the data above 340 MeV.}

\end{figure}
$\cal C$$_{\pi\pi}^A$ distributions divided by the reaction phase space, 
which also accounts for the CHAOS acceptance. In the $\pi^+\pi^+$ 
channel, $\cal C$$_{\pi\pi}^A$'s are flatly distributed at around 1, and 
the distributions cannot factually be distinguished from A=2 to A= 208. 
This clearly shows that the nuclear medium leaves the isospin 2 $\pi\pi$ 
interaction substantially unaltered. Strong medium modifications of the 
elementary $\pi\pi$ interaction are observed in the isospin 0 channel but 
only for $A>2$; in fact, the $\cal C$$_{\pi\pi}^A$ intensities display a 
sharp increase at around 2m$_\pi$ solely for $^{12}C, ^{40}Ca$ and 
$^{208}Pb$ (left panel). For $^2H$, $\cal C$$_{\pi\pi}^A$ is a relatively
flat distribution in  M$_{\pi\pi}$, without any indication of a threshold 
enhancement. 

For the examined nuclei, the composite ratio $\cal C$$_{\pi\pi}^A$  is 
compared to the model predictions 
\cite{theory:one,theory:two,theory:three}, which are normalized to the 
data at M$_{\pi\pi}>$340 MeV. The open diagram represents the results 
of Ref.\cite{theory:three} Fig. 7(a), for $\rho=0.5\rho_0$. In this 
model, the threshold enhancement is due to collective P-wave pionic 
modes as well 
as S-wave pionic modes, which largely contributes to the intensity 
at around the 2m$_\pi$ threshold. The  P-wave modes are described by 
the standard phenomenology of nuclear physics; that is, the $P-$wave 
coupling of pions to $particle-hole$ and $\Delta-hole$ configurations. 
The collective S-wave modes are studied via the Linear $\sigma$ Model 
developed to account for finite nuclear densities; in this study, the 
parameter $\rho$ can vary from 0 to 2$\rho_0$. The theory relates the 
$\pi\pi$ interaction in the scalar-isoscalar channel to the appearance 
of the $\sigma$-meson and,  finally, to the partial restoration of 
chiral symmetry in the nuclear medium. The filled diagram in Fig. 5 is
an earlier theoretical result \cite{theory:one}, which is based on the 
existence of the $\sigma$-meson in nuclear matter ($\rho=\rho_0$), 
$\sigma$ being generated by the fluctuation of the  $<\bar{q}q>$ chiral 
order parameter of QCD. The calculations depend on a complex parameter 
$\Phi(\rho)$, which for $\rho=\rho_0$ can vary from 0.7 to 0.9. In Fig. 
5, $\Phi(\rho=\rho_0)$=0.8. In the theory, the $\sigma$-meson reflects 
its existence by means of a marked enhancement of the spectral function 
at $\sim 2m_\pi$, which is a phenomenon commonly associated to the 
(partial) restoration of the chiral symmetry.  

The above comparison between experimental results and theoretical 
predictions clearly indicates that the $\pi\pi$ interaction is strongly 
modified by the presence of nuclei solely for I=J=0 pion pairs.
Furthermore, collective P-wave pion modes are far from  explaining the 
strength at $\sim 2m_\pi$; i.e., predictions of \cite{theory:four}
reported in Fig. 3. On the other hand, the inclusion  in the models of 
collective S-wave pionic modes is able to yield the requested threshold 
intensity even at $\rho < \rho_0$; i.e., calculations of Refs.
\cite{theory:one,theory:two,theory:three} reported in Fig. 5. 

\begin{center}
{\bf V. CONCLUSIONS}
\end{center}

The data discussed in the present letter are the results of an extended 
campaign of measurements of the $\pi N(A)\rightarrow \pi\pi N(A^\prime)$ 
reactions at several intermediate energies, which  involved the CHAOS 
collaboration at TRIUMF. Only charged  pions were detected, which 
allowed the $\pi\pi$ system to be studied in the isospin 0 and 2 
channels. The simultaneous study of the two isospin channels was 
essential to establish the correct size of the $\sigma$-strength. The 
only observable employed to reduce the data was the composite ratio 
$\cal C$$_{\pi\pi}^A$, which accounts for the clear effect of the 
nuclear medium on final pion pairs. Such an observable is nearly 
independent of the detector acceptance. 

In general, the nuclear medium has a negligible effect on the 
$(\pi\pi)_{I=2,J=0}$ system: the strength of the $\pi\pi$ interaction 
appears nearly the same in the vacuum as well as in the nuclear medium. 
In fact, the $\cal C$$_{\pi\pi}^A$ distributions are  planar at the
varying of M$_{\pi\pi}$ regardless of A. In the $\pi^+\pi^-$ channel, the 
$(\pi\pi)_{I=J=0}$ interaction is strongly modified by the medium even 
at moderate densities (i.e., $\rho\sim\frac{1}{3}\rho_0$), except for 
A=2 when nuclear matter is not realized. The distinctive signature of 
medium modification is the $\cal C$$_{\pi\pi}^A$ threshold enhancement.
These conclusions are common to all the examined energies, from 243 to
305 MeV. The $\cal C$$_{\pi\pi}^A$ distributions are finally compared to 
the theoretical predictions, which denote that the $\cal C$$_{\pi\pi}^A$ 
behavior is consistent with the effects of partial restoration of chiral 
symmetry in nuclear matter. Standard nuclear effects (such as Pauli 
blocking, Fermi motion, $\pi$N final-state interactions, etc.), or 
collective P-wave pionic modes are far from being able to explain the 
threshold enhancement.   

The threshold enhancement found for $\cal C$$_{\pi^+\pi^-}^A$ in the 
$\pi \rightarrow \pi\pi$ reaction \cite{expt:three,expt:four} was 
also found in the $\gamma \rightarrow \pi\pi$  reaction by the TAPS 
collaboration \cite{expt:six}. In the latter case, gammas  penetrate 
deeper into the nucleus thus being able to probe a nuclear density 
$\sim\frac{2}{3}\rho_0$ for $^{208}Pb$, which nearly doubles the 
density probed by the CHAOS measurements. Fig. 2 of Ref.\cite{expt:six} 
shows the TAPS $\cal C$$_{\pi\pi}$ ratio in comparison with the CHAOS 
ratio; in this case, the lead data are scaled to the carbon data. 
At the 2m$_\pi$ threshold, the TAPS ratio exceeds by $\sim$1.2 times 
the CHAOS ratio, which only partially reflects the higher density 
inspected by the TAPS measurements. On the other hand, Fig. 5 (left 
panel) yields $\cal C$$_{\pi\pi}^A\sim 8$ at threshold for C or Ca, 
which tends to favor the picture of a rapidly raising  $\sigma$ 
formation  as soon as an isospin 0 $\pi\pi$ system establishes
into the nuclear medium even of moderate density.

 It is finally worthwhile commenting the results of a recent 
theoretical work on the $\gamma\rightarrow\pi\pi$ reaction in 
nuclei\cite{theory:twelve}. The model examines the production 
and propagation of pion pairs in nuclear matter by using a 
semi-classical approximation and by fully accounting for the 
final state interactions of pions with the nuclear medium. The 
latter are found to distort considerably the $\pi \pi$ invariant 
mass distributions, which are then used for comparison to the 
TAPS data (i.e., Fig. 4 of Ref.\cite{theory:twelve}). The model 
predictions are capable of describing the threshold behavior of 
the I=0 $\gamma \rightarrow \pi^0\pi^0$ reaction channel for both 
$^{12}C$ and  $^{208}Pb$. The predictions from the same model 
however fail to reproduce the invariant mass distributions of 
the I=1 $\gamma \rightarrow \pi^0\pi^{+,-}$ channel. In fact, 
the curves show intensities 2-3 times higher than the data, and 
are 20-30 MeV downward peaked with respect to the experimental 
distributions. Regardless of the $\pi\pi$ reaction channel,
the mass distributions of Ref.\cite{theory:twelve} depict nearly 
the same threshold behavior, exactly where the effects of 
medium modification are experimentally observed. In order to 
account for such effects on $\pi\pi$ data, a model calculation 
must {\em simultaneously} explain the threshold behavior of the 
I=0 ($\sigma$) distributions as well as the I=1 (non-resonant) 
distributions. In this regard, the TAPS results are better 
predicted by the {\em Valencia} model (Fig. 4 of Ref.
\cite{theory:twelve}), which relates the peak shift of the 
$\pi\pi$ invariant mass distributions to the in-medium 
modification of the  $\pi\pi$ correlation. A general conclusion 
of this article is that to clearly probe the effects of chiral 
symmetry restoration, it is advisable to use the composite ratio 
$\cal C$$_{\pi\pi}^A$; in fact, this observable yields macroscopic 
threshold effects when comparing I=0 to I=1 or 2 isospin states, 
and it is nearly independent of the detector acceptance.   
 
\begin{center}
{\bf ACKNOWLEDGMENTS}
\end{center}

The present work was supported by the Istituto Nazionale di Fisica 
Nucleare (INFN) of Italy, the National Science and Engineering Research 
Council (NSERC) of Canada, the Australian Research Council and the 
German Ministry of Education and Research. The authors acknowledge the 
support received from TRIUMF. The authors would also like to acknowledge 
useful discussions with T. Kunihiro, P. Schuck and M. Vicente-Vacas.

%
%
 
%
%
\end{document}